\documentclass[12pt,a4paper]{article}

\usepackage{ifthen} 

\newboolean{pdflatex}
\setboolean{pdflatex}{true} 

\newboolean{articletitles}
\setboolean{articletitles}{true} 

\newboolean{uprightparticles}
\setboolean{uprightparticles}{false} 

\def\paperauthors{LHCb collaboration} 
\def\paperasciititle{Search for D0 meson decays to pipiee and kkee final states} 
\def\papertitle{Search for \Dz meson decays to $\pip\pim\ep\en$ and $\Kp\Km\ep\en$ final states} 
\def\paperkeywords{{High Energy Physics}, {LHCb}} 
\def\papercopyright{\the\year\ CERN for the benefit of the LHCb collaboration} 
\def\paperlicence{CC BY 4.0 licence}
\def\paperlicenceurl{https://creativecommons.org/licenses/by/4.0/}

\newif\ifEnableSectionTOCLinks
\EnableSectionTOCLinksfalse 

\usepackage[top=1in, bottom=1.25in, left=1in, right=1in]{geometry}

\usepackage{microtype}
\usepackage{lineno}  
\usepackage{xspace} 
\usepackage{caption} 

\usepackage{graphicx}  
\usepackage{color}
\usepackage{colortbl}
\graphicspath{{./figs/}} 

\usepackage{amsmath} 
\usepackage{amssymb}
\usepackage{amsfonts}
\usepackage{upgreek} 

\newcommand*\patchAmsMathEnvironmentForLineno[1]{%
\expandafter\let\csname old#1\expandafter\endcsname\csname #1\endcsname
\expandafter\let\csname oldend#1\expandafter\endcsname\csname
end#1\endcsname
 \renewenvironment{#1}%
   {\linenomath\csname old#1\endcsname}%
   {\csname oldend#1\endcsname\endlinenomath}%
}
\newcommand*\patchBothAmsMathEnvironmentsForLineno[1]{%
  \patchAmsMathEnvironmentForLineno{#1}%
  \patchAmsMathEnvironmentForLineno{#1*}%
}
\AtBeginDocument{%
\patchBothAmsMathEnvironmentsForLineno{equation}%
\patchBothAmsMathEnvironmentsForLineno{align}%
\patchBothAmsMathEnvironmentsForLineno{flalign}%
\patchBothAmsMathEnvironmentsForLineno{alignat}%
\patchBothAmsMathEnvironmentsForLineno{gather}%
\patchBothAmsMathEnvironmentsForLineno{multline}%
\patchBothAmsMathEnvironmentsForLineno{eqnarray}%
}

\usepackage{hyperref}
\usepackage{hyperxmp}
\hypersetup{pdftex,
pdfauthor={\paperauthors},
pdftitle={\paperasciititle},
pdfkeywords={\paperkeywords},
pdfcopyright={Copyright (C) \papercopyright},
pdflicenseurl={\paperlicenceurl}}

\usepackage[colorinlistoftodos,textsize=scriptsize]{todonotes}

\usepackage[bottom,flushmargin,hang,multiple]{footmisc}

\usepackage[all]{hypcap} 

\usepackage{xspace} 
\usepackage{upgreek}

\def\lhcb   {\mbox{LHCb}\xspace}

\def\MagUp {\mbox{\em Mag\kern -0.05em Up}\xspace}

\ifthenelse{\boolean{uprightparticles}}%
{

 \def\Ppi         {\ensuremath{\uppi}\xspace}                 
                  
 \def\Prho        {\ensuremath{\uprho}\xspace}

 \def\Pphi        {\ensuremath{\upphi}\xspace}

 \def\Pomega      {\ensuremath{\upomega}\xspace}                 

 \def\PDelta      {\ensuremath{\Delta}\xspace}                 
 \def\PXi         {\ensuremath{\Xi}\xspace}                 
 \def\PLambda     {\ensuremath{\Lambda}\xspace}                 
 \def\PSigma      {\ensuremath{\Sigma}\xspace}                 
 \def\POmega      {\ensuremath{\Omega}\xspace}                 
 \def\PUpsilon    {\ensuremath{\Upsilon}\xspace}
 \let\oldPi\Pi
 \def\PPi         {\ensuremath{\oldPi}\xspace}

 \def\PB      {\ensuremath{\mathrm{B}}\xspace}                 
 \def\PD      {\ensuremath{\mathrm{D}}\xspace}                 

 \def\PK      {\ensuremath{\mathrm{K}}\xspace}                 

 \def\Pe      {\ensuremath{\mathrm{e}}\xspace}                 
 \def\Pp      {\ensuremath{\mathrm{p}}\xspace}                 

 \def\Ps      {\ensuremath{\mathrm{s}}\xspace}

 \def\thebaroffset{0.0em}
}
{

 \def\Ppi         {\ensuremath{\pi}\xspace}                 
                  
 \def\Prho        {\ensuremath{\rho}\xspace}

 \def\Pphi        {\ensuremath{\phi}\xspace}

 \def\Pomega      {\ensuremath{\omega}\xspace}                 
 \mathchardef\PDelta="7101
 \mathchardef\PXi="7104
 \mathchardef\PLambda="7103
 \mathchardef\PSigma="7106
 \mathchardef\POmega="710A
 \mathchardef\PUpsilon="7107
 \mathchardef\PPi="7105
 \def\PB      {\ensuremath{B}\xspace}                 
 \def\PD      {\ensuremath{D}\xspace}                 

 \def\PK      {\ensuremath{K}\xspace}                 

 \def\Pe      {\ensuremath{e}\xspace}                 
 \def\Pp      {\ensuremath{p}\xspace}                 

 \def\Ps      {\ensuremath{s}\xspace}

 \def\thebaroffset{0.18em}
}
\newcommand{\offsetoverline}[2][\thebaroffset]{\kern #1\overline{\kern -#1 #2}}%

\makeatletter
\ifcase \@ptsize \relax
  \newcommand{\miniscule}{\@setfontsize\miniscule{4}{5}}
\or
  \newcommand{\miniscule}{\@setfontsize\miniscule{5}{6}}
\or
  \newcommand{\miniscule}{\@setfontsize\miniscule{5}{6}}
\fi
\makeatother

\DeclareRobustCommand{\optbar}[1]{\shortstack{{\miniscule (\rule[.5ex]{1.25em}{.18mm})}
  \\ [-.7ex] $#1$}}

\def\en         {{\ensuremath{\Pe^-}}\xspace}   
\def\ep         {{\ensuremath{\Pe^+}}\xspace}

\def\epem       {{\ensuremath{\Pe^+\Pe^-}}\xspace}

\def\ellell     {\ensuremath{\ell^+ \ell^-}\xspace}

\def\squark    {{\ensuremath{\Ps}}\xspace}

\def\pion   {{\ensuremath{\Ppi}}\xspace}

\def\pip    {{\ensuremath{\pion^+}}\xspace}
\def\pim    {{\ensuremath{\pion^-}}\xspace}

\def\rhomeson {{\ensuremath{\Prho}}\xspace}
\def\rhoz     {{\ensuremath{\rhomeson^0}}\xspace}

\def\kaon    {{\ensuremath{\PK}}\xspace}

\def\KorKbar {\kern \thebaroffset\optbar{\kern -\thebaroffset \PK}{}\xspace}

\def\Kp      {{\ensuremath{\kaon^+}}\xspace}
\def\Km      {{\ensuremath{\kaon^-}}\xspace}

\newcommand{\phiz}{\ensuremath{\Pphi}\xspace}
\newcommand{\omegaz}{\ensuremath{\Pomega}\xspace}

\def\D       {{\ensuremath{\PD}}\xspace}

\def\DorDbar {\kern \thebaroffset\optbar{\kern -\thebaroffset \PD}\xspace}
\def\Dz      {{\ensuremath{\D^0}}\xspace}

\def\Dp      {{\ensuremath{\D^+}}\xspace}
\def\Dm      {{\ensuremath{\D^-}}\xspace}

\def\DpDm    {\ensuremath{\Dp {\kern -0.16em \Dm}}\xspace}

\def\Dstarp  {{\ensuremath{\D^{*+}}}\xspace}

\def\B       {{\ensuremath{\PB}}\xspace}

\def\BorBbar {\kern \thebaroffset\optbar{\kern -\thebaroffset \PB}\xspace}

\def\Bd      {{\ensuremath{\B^0}}\xspace}

\def\BdorBdbar {\kern \thebaroffset\optbar{\kern -\thebaroffset \Bd}\xspace}

\def\Bs      {{\ensuremath{\B^0_\squark}}\xspace}

\def\BsorBsbar {\kern \thebaroffset\optbar{\kern -\thebaroffset \Bs}\xspace}

\def\Y#1S{\ensuremath{\PUpsilon{(#1S)}}\xspace}

\def\proton      {{\ensuremath{\Pp}}\xspace}

\def\LorLbar     {\kern \thebaroffset\optbar{\kern -\thebaroffset \PLambda}\xspace}

\def\BF         {{\ensuremath{\mathcal{B}}}\xspace}

\def\to                 {\ensuremath{\rightarrow}\xspace}

\newcommand{\dm}{{\ensuremath{\Delta m}}\xspace}

\def\AT#1     {\ensuremath{A_{\mathrm{T}}^{#1}}\xspace}           

\def\C#1      {\ensuremath{\mathcal{C}_{#1}}\xspace}                       
\def\Cp#1     {\ensuremath{\mathcal{C}_{#1}^{'}}\xspace}                    
\def\Ceff#1   {\ensuremath{\mathcal{C}_{#1}^{\mathrm{(eff)}}}\xspace}        
\def\Cpeff#1  {\ensuremath{\mathcal{C}_{#1}^{'\mathrm{(eff)}}}\xspace}       
\def\Ope#1    {\ensuremath{\mathcal{O}_{#1}}\xspace}                       
\def\Opep#1   {\ensuremath{\mathcal{O}_{#1}^{'}}\xspace}                    

\newcommand{\aunit}[1]{\ensuremath{\text{\,#1}}}       

\newcommand{\tev}{\aunit{Te\kern -0.1em V}\xspace}
\newcommand{\gev}{\aunit{Ge\kern -0.1em V}\xspace}
\newcommand{\mev}{\aunit{Me\kern -0.1em V}\xspace}
\newcommand{\kev}{\aunit{ke\kern -0.1em V}\xspace}
\newcommand{\ev}{\aunit{e\kern -0.1em V}\xspace}
 
\newcommand{\mevc}{\ensuremath{\aunit{Me\kern -0.1em V\!/}c}\xspace}
\newcommand{\gevc}{\ensuremath{\aunit{Ge\kern -0.1em V\!/}c}\xspace}
\newcommand{\mevcc}{\ensuremath{\aunit{Me\kern -0.1em V\!/}c^2}\xspace}
\newcommand{\gevcc}{\ensuremath{\aunit{Ge\kern -0.1em V\!/}c^2}\xspace}

\def\fb   {\ensuremath{\aunit{fb}}\xspace}
\def\invfb   {\ensuremath{\fb^{-1}}\xspace}

\def\gsim{{~\raise.15em\hbox{$>$}\kern-.85em
          \lower.35em\hbox{$\sim$}~}\xspace}
\def\lsim{{~\raise.15em\hbox{$<$}\kern-.85em
          \lower.35em\hbox{$\sim$}~}\xspace}

\def\pt         {\ensuremath{p_{\mathrm{T}}}\xspace}

\def\tell1  {TELL1\xspace}
\def\ukl1   {UKL1\xspace}

\newcommand{\eg}{\mbox{\itshape e.g.}\xspace}

\newcommand{\lhcborcid}[1]{\href{https://orcid.org/#1}{\hspace*{0.1em}\raisebox{-0.45ex}{\includegraphics[width=1em]{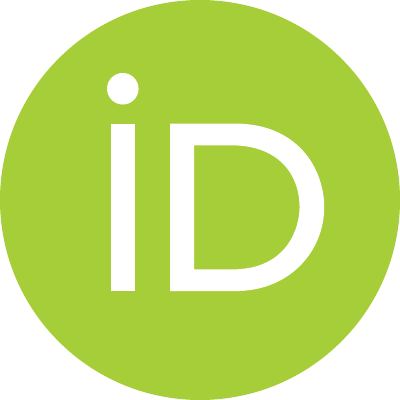}}}}

\hypersetup{
  colorlinks   = true, 
  urlcolor     = blue, 
  linkcolor    = blue, 
  citecolor    = red   
}

\ifEnableSectionTOCLinks
    \usepackage[explicit]{titlesec} 
    
    \let\oldcontentsline\contentsline
    \renewcommand

    \titleformat{\section}{\normalfont\Large\bf}{\hyperlink{tocsection.\thesection}{{\thesection} \parbox[t]{\dimexpr\textwidth-1pc}{#1}}}{1pc}{}

    \titleformat{\subsection}{\normalfont\bf}{\hyperlink{tocsubsection.\thesubsection}{{\thesubsection} \parbox[t]{\dimexpr\textwidth-1pc}{#1}}}{1pc}{}

    \titleformat{name=\section,numberless}[display]{}{}{0pt}{\normalfont\Huge\bfseries #1}
\fi

\usepackage{cite} 
\usepackage{mciteplus}

\usepackage{color}

\newcommand{\hhp}{\ensuremath{h^+h^{({\mkern-1mu\prime})-}}\xspace}

\newcommand{\hhee}{\ensuremath{h^+h^-\ep\en}\xspace}

\newcommand{\kpee}{\ensuremath{\Km\pip\ep\en}\xspace}

\newcommand{\kkee}{\ensuremath{\Kp\Km\ep\en}\xspace}
\newcommand{\ppee}{\ensuremath{\pip\pim\ep\en}\xspace}
\newcommand{\Dhhpmm}{\ensuremath{\Dz\to\hhp\mu^+\mu^-}\xspace}
\newcommand{\Dhhmm}{\ensuremath{\Dz\to h^+h^-\mu^+\mu^-}\xspace}
\newcommand{\Dhhee}{\ensuremath{\Dz\to h^+h^-\ep\en}\xspace}
\newcommand{\Dkkmm}{\ensuremath{\Dz\to\Kp\Km\mu^+\mu^-}\xspace}
\newcommand{\Dppmm}{\ensuremath{\Dz\to\pip\pim\mu^+\mu^-}\xspace}
\newcommand{\Dkkee}{\ensuremath{\Dz\to\Kp\Km\ep\en}\xspace}
\newcommand{\Dppee}{\ensuremath{\Dz\to\pip\pim\ep\en}\xspace}

\newcommand{\Dkpee}{\ensuremath{\Dz\to\Km\pip[\ep\en]_{\rhoz/\omegaz}}\xspace}
\newcommand{\Dkpeenr}{\ensuremath{\Dz\to\Km\pip\ep\en}\xspace}

\newcommand{\Dhhpee}{\ensuremath{\Dz\to\hhp\ep\en}\xspace}
\newcommand{\Dppeetot}{\ensuremath{\Dz\to\pip\pim[\ep\en]_{m(\ep\en)>2m_\mu}}\xspace}

\newcommand{\mD}{\ensuremath{m(\Dz)}\xspace}

\newboolean{wordcount}
\setboolean{wordcount}{false} 

\usepackage{longtable} 

\begin{document}

\renewcommand{\thefootnote}{\fnsymbol{footnote}}
\setcounter{footnote}{1}


\begin{titlepage}
\pagenumbering{roman}

\vspace*{-1.5cm}
\centerline{\large EUROPEAN ORGANIZATION FOR NUCLEAR RESEARCH (CERN)}
\vspace*{1.5cm}
\noindent
\begin{tabular*}{\linewidth}{lc@{\extracolsep{\fill}}r@{\extracolsep{0pt}}}
\ifthenelse{\boolean{pdflatex}}
{\vspace*{-1.5cm}\mbox{\!\!\!\includegraphics[width=.14\textwidth]{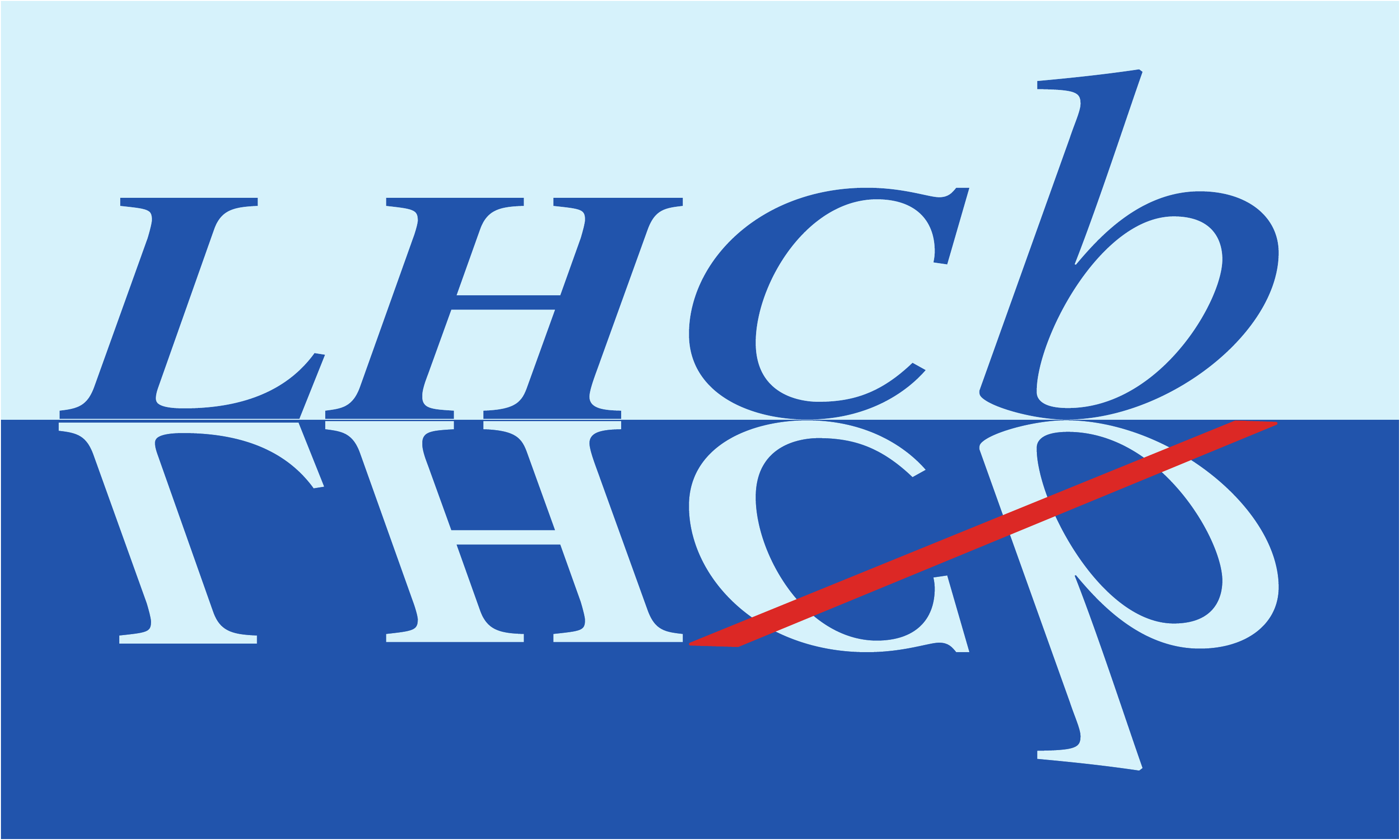}} & &}%
{\vspace*{-1.2cm}\mbox{\!\!\!\includegraphics[width=.12\textwidth]{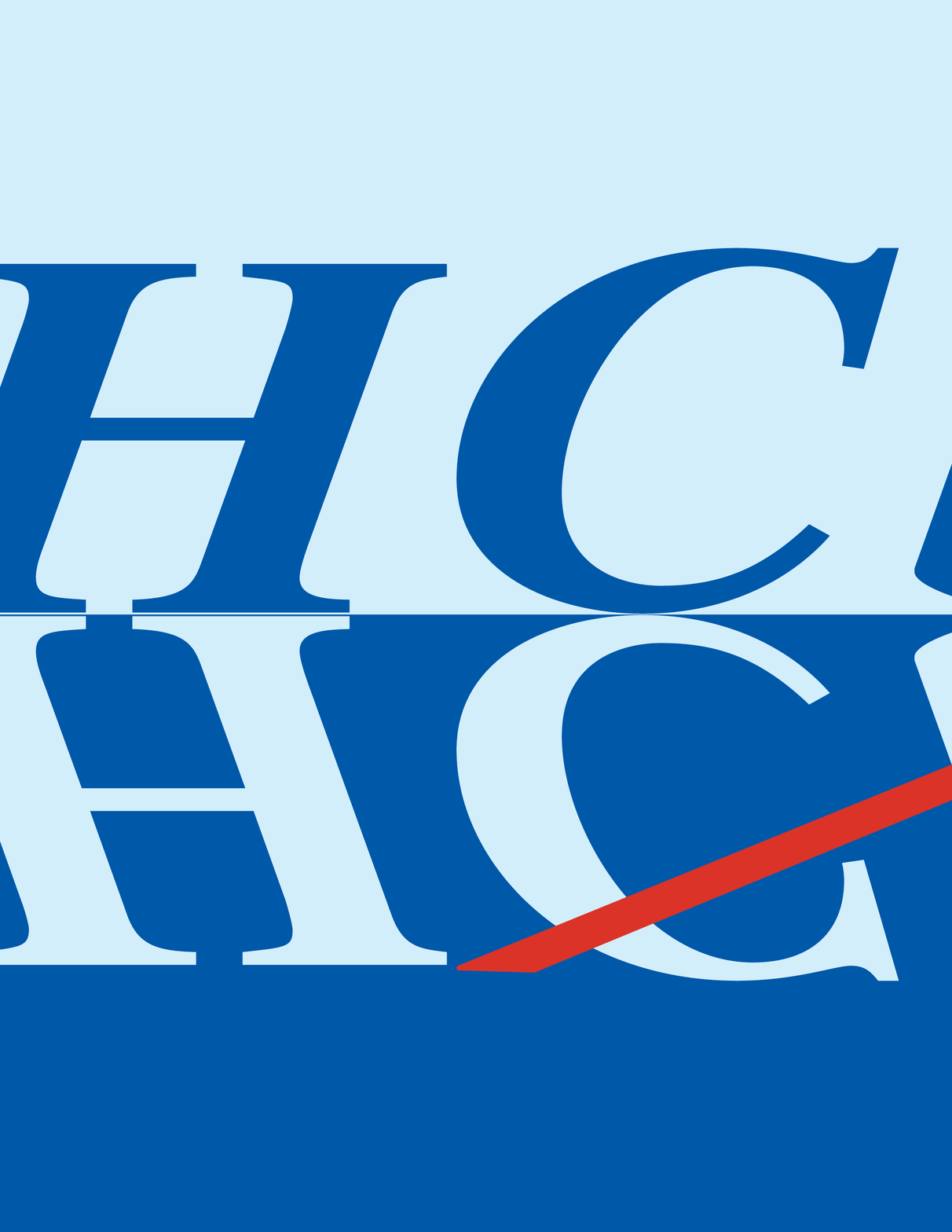}} & &}%
\\
 & & CERN-EP-2024-307 \\  
 & & LHCb-PAPER-2024-047 \\  
 & & May 5, 2025 \\ 
 & & \\
\end{tabular*}

\vspace*{3.0cm}

{\normalfont\bfseries\boldmath\huge
\begin{center}
  \papertitle 
\end{center}
}

\vspace*{2.0cm}

\begin{center}
\paperauthors\footnote{Authors are listed at the end of this paper.}
\end{center}

\vspace{\fill}

\begin{abstract}
  \noindent
  A search for $D^0$ meson decays to the $\pi^+\pi^-e^+e^-$ and $K^+K^-e^+e^-$ final states is reported using a sample of proton-proton collisions collected by the LHCb experiment at a center-of-mass energy of 13\,{Te\kern -0.1em V}, corresponding to an integrated luminosity of 6\,fb$^{-1}$. The decay $D^0 \rightarrow \pi^+\pi^-e^+e^-$ is observed for the first time when requiring that the two electrons are consistent with coming from the decay of a $\phi$ or $\rho^0/\omega$ meson. The corresponding branching fractions are measured relative to the \mbox{$D^0 \rightarrow K^-\pi^-[e^+e^-]_{\rho^0/\omega}$} decay, where the two electrons are consistent with coming from the decay of a $\rho^0$ or $\omega$ meson. 
  No evidence is found for the $D^0 \rightarrow K^+K^-e^+e^-$ decay and world-best limits are set on its branching fraction. The results are compared to, and found to be consistent with, the branching fractions of the $D^0 \rightarrow \pi^+\pi^-\mu^+\mu^-$ and $D^0 \rightarrow K^+K^-\mu^+\mu^-$ decays recently measured by LHCb and confirm lepton universality at the current precision.  
  
\end{abstract}

\vspace*{2.0cm}

\begin{center}
  Published in Phys. Rev. D111 (2025) L091101
\end{center}

\vspace{\fill}

{\footnotesize 
\centerline{\copyright~\papercopyright. \href{\paperlicenceurl}{\paperlicence}.}}
\vspace*{2mm}

\end{titlepage}


\newpage
\setcounter{page}{2}
\mbox{~}
%
%
%
%


\renewcommand{\thefootnote}{\arabic{footnote}}
\setcounter{footnote}{0}


\cleardoublepage


\pagestyle{plain} 
\setcounter{page}{1}
\pagenumbering{arabic}


Within the standard model (SM), flavor-changing neutral-current (FCNC) processes are suppressed by the Glashow--Iliopoulos--Maiani mechanism~\cite{GIM}. Extensions of the SM can, however, 
significantly alter the probabilities at which these processes occur. Depending on the nature of potential beyond-SM (BSM) contributions, measurements of observables sensitive to FCNCs can probe energy scales of tens or even hundreds of~TeV\cite{EuropeanStrategyforParticlePhysicsPreparatoryGroup:2019qin},
providing a powerful tool in characterizing the allowed parameter space of BSM physics. 

Rare charm decays may proceed via FCNC $c\to u\ell^+\ell^-$ transitions ($\ell=\mu$ or $e$). In the SM, these so-called {\em short-distance} contributions result in branching fractions of $\mathcal{O}(10^{-9})$~\cite{PaulBigi:2011} 
for decays of the type $\D\to X\ell^+\ell^-$, where $X$ is one or more hadrons. The experimentally observable $\D\to X\ell^+\ell^-$ decays are, however, dominated by {\em long-distance} processes involving intermediate hadronic resonances such as \mbox{$\D\to XY(\to\ell^+\ell^-)$} where $Y$ is a hadronic resonance, \eg a 
\rhoz, \omegaz or \phiz meson. These resonances enhance the SM branching fractions but remain suppressed at the level of $\mathcal{O}(10^{-6})$~\cite{Fajfer:2007,PaulBigi:2011,Cappiello,deBoer:2018}, with the broadest resonances spanning the entire dilepton-mass spectrum. 
Accessing the short-distance contributions of interest therefore requires both large data sets and the use of complementary observables that are sensitive to short- and long-distance processes, such as angular distributions, charge-parity asymmetries or tests of lepton universality~\cite{Fajfer:2005ke,Bigi:2012,Paul:2012ab,Fajfer:2012nr,Cappiello,Fajfer:2015mia,deBoer:2015boa,deBoer:2018,Gisbert:2020vjx,Bause:2019vpr,Bause:2020obd,Bharucha:2020eup,Fajfer:2023tkp,Mandal:2024lbp,Gisbert:2024kob}. 

Due to the universality of the electroweak interaction couplings to leptons in the SM, any differences in meson decay rates into final states with leptons of different generations arise purely from lepton mass effects.
Consequently, comparing the branching fraction of a decay mode with muons in the final state to its counterpart with electrons provides a powerful probe of BSM effects, despite the dominance of long-distance contributions and large uncertainties in the SM prediction for the individual decay modes. Depending on the properties of possible BSM contributions in $c\to u\ell^+\ell^-$ transitions, the validity of lepton universality could be broken at the percent level~\cite{deBoer:2018}. 

The \Dhhmm decays ($h=\pi$ or $K$) were first observed by the LHCb collaboration~\cite{LHCb-PAPER-2017-019}, which subsequently studied their angular structure and charge-parity asymmetries~\cite{LHCb-PAPER-2018-020,LHCb-PAPER-2021-035}. Charge-conjugate decays are implied throughout the Letter. The \Dhhee decays offer the opportunity to probe the validity of lepton universality.
However, electron final states are more challenging experimentally, in particular because the electrons lose a significant fraction of their energy as bremsstrahlung radiation while passing through the detector material. Until now, only upper limits on the branching fractions of \Dhhee decays have been reported \cite{E791:2000jkj,BESIII:2018hqu}.

This Letter reports the first search for the \Dppee and \Dkkee decays at the \lhcb experiment and  opens the door for the first lepton universality tests with such decays. The analysis uses $\proton\proton$ collision data collected between 2015 and 2018 at a center-of-mass energy $\sqrt{s}=13\tev$, corresponding to an integrated luminosity of 6\invfb. 
The analysis method closely follows that of the LHCb analysis which observed the analogous decays with muons~\cite{LHCb-PAPER-2017-019}, however the dielectron final state requires the development of a specific selection and reconstruction. The analysis uses \Dz mesons originating from $\Dstarp\to\Dz\pip$ decays with the $\Dstarp$ mesons produced in the primary $pp$ interaction. 

The branching fractions of the decays of interest are measured in regions of the dielectron mass relative to the \Dkpee decay which has a branching fraction of \mbox{$(4.0 \pm 0.5) \times 10^{-6}$} in the dielectron-mass range $675$--$875\mevcc$~\cite{exp:BABAR}, where the notation indicates that the contribution from the $\rhoz/\omegaz\to\ep\en$ decay is dominant. 

The \lhcb detector is a single-arm forward spectrometer covering the pseudorapidity
range $2 <\eta< 5$ and is described in detail in Refs.~\cite{LHCb-DP-2008-001,LHCb-DP-2014-002}.
Events are selected by a trigger that consists of a hardware stage, based on information from the calorimeter and muon systems, followed by a software stage which applies a full event reconstruction~\cite{LHCb-DP-2012-004}.
The hardware trigger requires the presence of an energy deposit in the electromagnetic calorimeter that may or may not be compatible with originating from the signal candidate, or a muon signature with large transverse momentum which is compatible with originating from any particle in the event.
A first stage of the software trigger selects events with either a charged particle that has both a significant transverse momentum and large impact parameter, defined as the minimum distance of the particle trajectory from any primary \proton\proton-collision vertex (PV), or with a two-track vertex satisfying a multivariate classifier based on geometric and kinematic criteria which identify the vertex as likely to originate from the decay of a long-lived heavy particle. 
In a second stage of the software trigger, candidate \Dhhpee decays are selected by combining four tracks that form a secondary vertex separated from any PV. All charged particles are required to have a significant impact parameter, as well as momentum $p>3\gevc$ and transverse momentum $\pt>0.3\gevc$. 
The \Dz candidate must have large transverse momentum, a reconstructed mass, $m(\Dz)$, in the range 1800--1950\mevcc (1700--2050\mevcc) for data recorded in the years 2015--16 (2017--18) and its momentum vector must be aligned with the vector connecting the PV and the \Dz decay vertex. The larger mass range in the second period is due to a change in the software trigger that allows for improved background studies using information from the sidebands. When more than one PV is reconstructed, the one with respect to which the \Dz candidate has the lowest impact-parameter $\chi^2$ is chosen, defined as
the difference in the vertex-fit $\chi^2$ reconstructed with
and without the candidate. Finally, $\Dstarp\to\Dz\pip$ candidates are selected by combining the \Dz meson with a charged particle from the same PV that has $\pt>120\mevc$. 

Subsequently, signal candidates are further selected offline by tightening the kinematic and geometric criteria applied in the trigger. A dedicated algorithm associates reconstructed bremsstrahlung photons to tracks identified as electrons; when a given photon is associated with both electron tracks, it is attached to one chosen randomly.
Throughout the analysis, signal candidates are split into two periods with different data-taking conditions (2015--16 and 2017--18) and further divided into two categories: candidates in which neither electron has an associated bremsstrahlung cluster, and all other candidates, referred to as the no-brem and with-brem categories, respectively. A multivariate classifier is used to remove \Dz candidates containing one or more fake tracks. Stringent particle-identification criteria are then applied to all charged particles to suppress both combinatorial background, from unrelated charged particles, and cross-feed backgrounds in which one type of \Dhhpee decay is misidentified as another. 
The vertex formed by the \Dz and \pip mesons is constrained to coincide with the PV and the momenta of the particles in the decay chain are updated accordingly. Only candidates with a difference between the reconstructed \Dstarp and \Dz masses, \dm, in the range $144$--$147$\mevcc are considered.
 Further reduction of combinatorial background is achieved using a multivariate selection based on a boosted decision tree (BDT)~\cite{Breiman,Roe}.
The following features are used to discriminate signal from background: the momentum, transverse momentum, and impact parameter of the pion from the \Dstarp decay; the fit quality of the \Dz vertex and its separation from the PV; the angle between the \Dz momentum vector and the vector connecting the PV and the \Dz decay vertex; and the fit quality of the \Dstarp vertex. The BDT classifier is trained separately for \Dppee and \Dkkee decays, using simulation samples~\cite{LHCb-PROC-2011-006, LHCb-PROC-2010-056} as a proxy for the signal and data candidates with \mD greater than 1900\mevcc as a proxy for the background. 

Hadronic $\Dz\to\pip\pim\pip\pim$ and $\Dz\to\Kp\Km\pip\pim$ decays, where two pions are misidentified as electrons, constitute a major source of background, which is reduced by a multivariate electron-identification discriminant that combines information from the Cherenkov detectors, the calorimeters and the muon chambers.
Finally, selection criteria based on the BDT response and on the electron identification discriminant are optimized by maximizing the figure-of-merit ${\epsilon_\text{\hhee}}/({5/2 + \sqrt{N_\text{bkg}}})$~\cite{Punzi:2003bu}, where $\epsilon_\text{\hhee}$ is the signal efficiency, and $N_\text{bkg}$ is the total background yield in the \mD range $1700$--$1900$\mevcc ($1700$--$2050$\mevcc) for the no-brem (with-brem) category. 
The no-brem category has a significantly larger contamination from backgrounds in which hadrons are wrongly identified as electrons and therefore significantly worse sensitivity to the signals of interest. 
To facilitate an extrapolation of the yields of misidentified background from the no-brem to the with-brem category, the optimal selection thresholds obtained for the with-brem category are used for the no-brem category.
Candidate \Dkpee decays are selected using the response of the BDT classifier trained on the \Dppee (\Dkkee) signal when being used to normalize the branching fraction of \Dppee (\Dkkee). Only one candidate is kept at random if an event contains several signal or normalization candidate decays after the final selection, which happens in less than 0.5\% of selected events. To avoid potential biases on the measured signal candidate yields, candidate decays in the \mD signal region were examined only after the analysis procedure was finalized.

\begin{figure}[tb]
\centering
\includegraphics[width=0.465\textwidth]{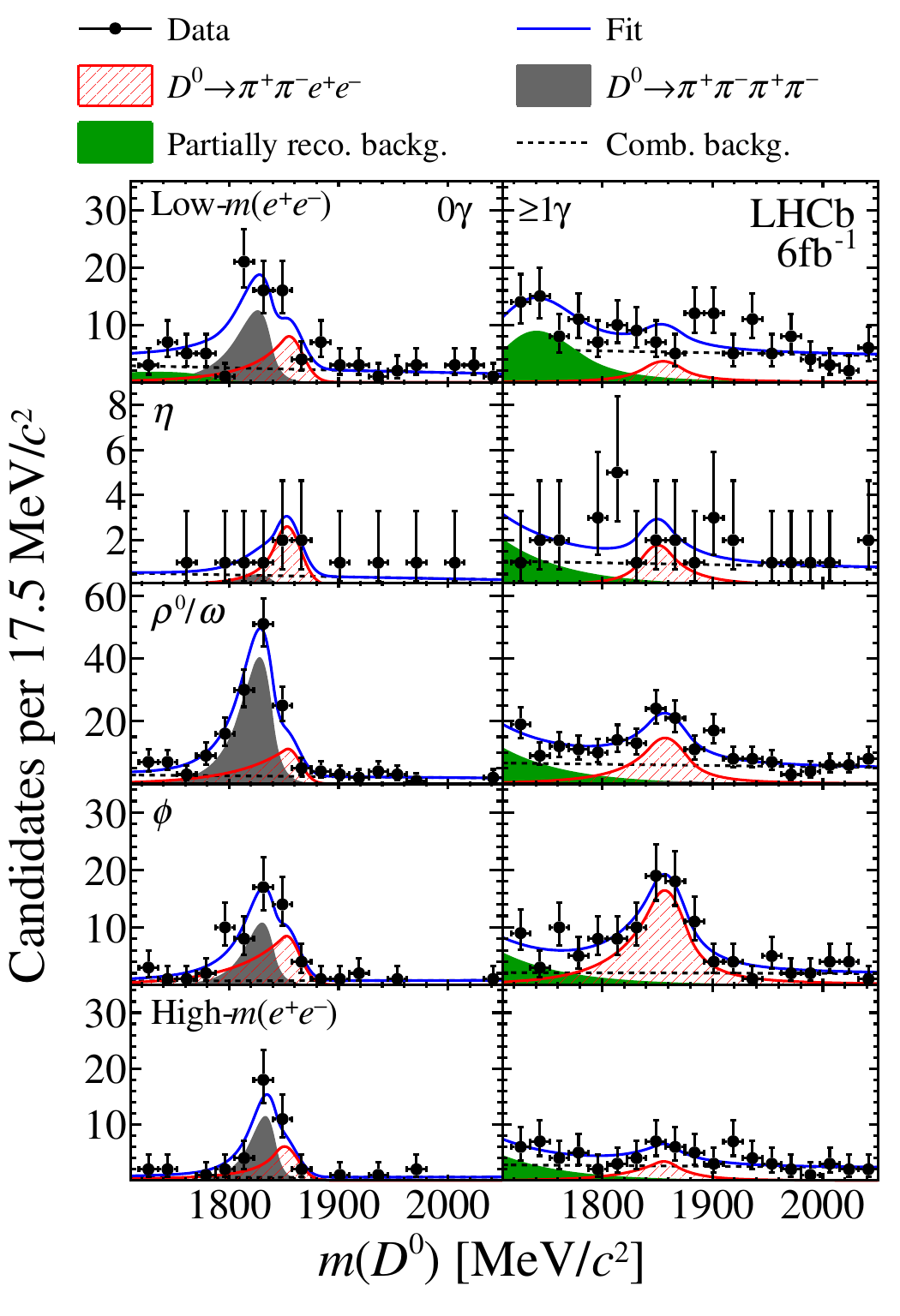}
\caption{Mass distributions of selected $D^0 \rightarrow \pi^+\pi^-e^+e^-$ candidates in the low-$m(e^+e^-)$, $\eta$, $\rho^0/\omega$, $\phi$ and high-$m(e^+e^-)$ regions in the (left, $0\gamma$) no-brem and (right, $\geq 1\gamma$) with-brem  categories. Fit projections are also shown.\label{fig:massfits-ppmm}}
\end{figure}

\begin{figure}[tb]
\centering
\includegraphics[width=0.465\textwidth]{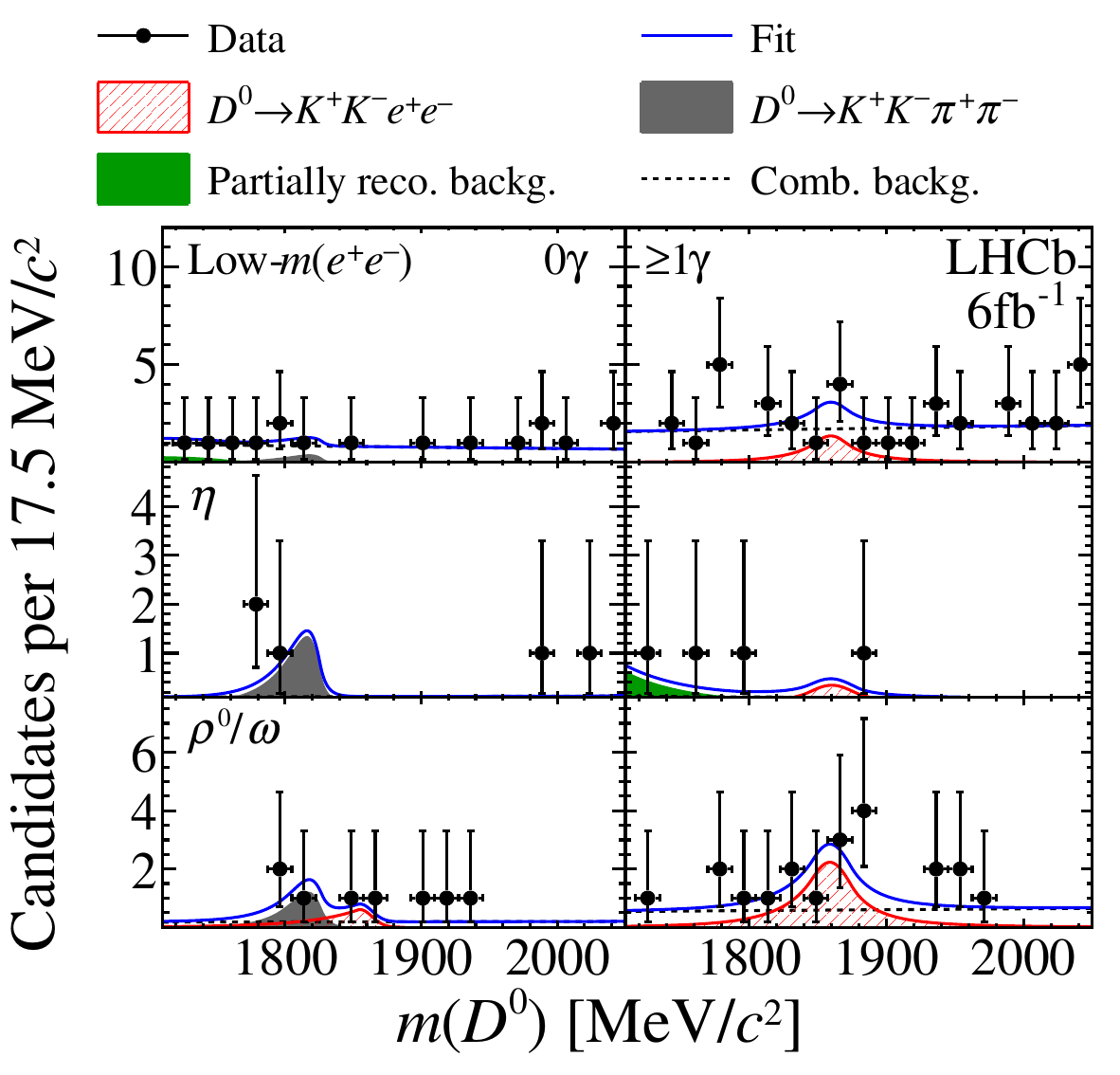}
\caption{Mass distributions of selected $D^0 \rightarrow K^+K^-e^+e^-$ candidates in the low-$m(e^+e^-)$, $\eta$ and $\rho^0/\omega$ regions in the (left) no-brem ($0\gamma$) and (right) with-brem ($\geq 1\gamma$) categories. Fit projections are also shown. \label{fig:massfits-kkmm}}
\end{figure}

Signal and normalization yields are measured with unbinned extended maximum-likelihood fits to the \mD distributions, in regions of the dielectron mass, $m(\ep\en)$, which is computed constraining the mass of the \Dz meson to its known value\cite{PDG2024}. The fit results are shown for \Dppee and \Dkkee candidates in Figs.~\ref{fig:massfits-ppmm} and \ref{fig:massfits-kkmm}, respectively. The fit results for \Dkpee can be found in the supplemental material of this letter~\cite{supplemental}.
For the signal modes, the $m(\ep\en)$ regions are defined in Table~\ref{tab:yields-effs} according to the presence of known intermediate resonances and aligned with Ref.~\cite{LHCb-PAPER-2018-020}. Thus, the lowest $m(e^+e^-)$ region starts at two times the muon mass, $m_\mu$.
All fits include four main components: signal, combinatorial background, peaking background from misidentified hadronic decays and background from partially reconstructed $\Dz$-meson decays, where at least one charged or neutral particle has not been reconstructed. The fit also accounts for the small fraction of $\Dz\to h^+h^-\ep\en$ decays in which the dielectron pair is produced in one $m(e^+e^-)$ region and reconstructed in another. 
All fits are validated to return unbiased results using large numbers of pseudoexperiments.
The signal is described with a Crystal Ball~\cite{Skwarnicki:1986xj} distribution for the no-brem category and with a double-sided Crystal Ball distribution for the with-brem category. In both cases, the distribution parameters are determined from simulation.
The mass shape of the peaking background is determined using a Bukin distribution~\cite{bukin2007fittingfunctionasymmetricpeaks} fitted to simulated samples of $\Dz\to h^+h^{(')-}\pip\pim$ decays, where the \Dz mass is calculated by assigning both pion candidates the electron mass. The peaking background in the with-brem category has a small, but not negligible, yield and a shape which is very similar to the signal. Its yield in the with-brem category is extrapolated from the no-brem category using the relevant efficiencies from simulation.
The combinatorial background is described by a first-order polynomial function with its slope determined from data candidates with $\dm > 150\mevcc$ and $ \mD > 1900\mevcc$. All shape parameters of the signal, peaking background and combinatorial background are fixed. Partially reconstructed backgrounds are modeled using a Bukin distribution fitted to simulated events in the low $m(e^+e^-)$ region ($<525\mevcc$), and an exponential function with its shape parameter determined in the fit to the selected candidates in other $m(e^+e^-)$ ranges. Alternative parametrizations are studied as a source of systematic uncertainty.
The yields of each component are allowed to vary in the fits, which are performed simultaneously between the bremsstrahlung categories, periods of data-taking, and dielectron-mass regions. 
The yield of the combinatorial background is constrained to the yield determined in the \dm sideband $155$--$165\mevcc$ extrapolated to the signal region.

\begin{table}[tb]
\centering
\caption{Yields of (top) \Dppee and (bottom) \Dkkee signal decays and their significance, $\mathcal{S}$, in units of Gaussian standard deviations, with respect to zero.}\label{tab:yields-effs}
\ifthenelse{\boolean{wordcount}}{}{%
\begin{tabular}{lr@{--}lr@{\,$\pm$\,}lr} 
$m(\epem)$ region & \multicolumn{2}{c}{[$\mevcc$]} & \multicolumn{2}{c}{Yield} & \multicolumn{1}{c}{$\mathcal{S}$}  \\
\hline
\multicolumn{6}{c}{\Dppee}\\
Low mass        & 2$m_\mu$ & 525  &  37 &  13 & $2.8\sigma$ \\
$\eta$          & 525 & 565                   &   10 &  7 & $1.6\sigma$  \\
$\rhoz/\omegaz$ & 565 & 950                   & 97 & 21 &  $5.5\sigma$  \\
$\phi$          & 950 & 1100                  & 100 & 18 &  $8.1\sigma$  \\
High mass       & \multicolumn{2}{c}{$>1100$} &   30 &  11 & $2.9\sigma$  \\
\hline
\multicolumn{6}{c}{\Dkkee}\\
Low mass        & 2$m_\mu$ & 525 &  4 &  8 & $1.2\sigma$  \\
$\eta$          & 525 & 565                   & 1 & 2  & $1.1\sigma$  \\
$\rhoz/\omegaz$ & \multicolumn{2}{c}{$>565$}  & 12 &  7 & $2.2\sigma$ \\
\hline
\end{tabular}}
\end{table}

The resulting signal yields and significances with respect to zero, including statistical and systematic uncertainties, are reported in Table~\ref{tab:yields-effs}. Significances exceeding five standard deviations are reported for \Dppee decays, where the two electrons are consistent with coming from an intermediate $\phiz$ or $\rhoz/\omega$ meson.

The signal yields, $N^i_{\hhee}$, in the $m(\ep\en)$ region $i$ are used to compute the branching fractions as
\ifthenelse{\boolean{wordcount}}{}{%
\begin{equation*}
\BF^i(\Dhhee) = \frac{N_{\hhee}^i\, \BF(\Dkpee)}{R^i_\epsilon\,N_{\kpee}},
\end{equation*}
}
where $N_{\kpee}$ is the yield of the normalization mode, which is determined to be \mbox{$820\pm 39$} ($875\pm 40$) after applying the selection optimized for \mbox{\Dppee} (\mbox{\Dkkee}) decays, while \mbox{$R_\epsilon^i=\epsilon_{\hhee}^i/\epsilon_{\kpee}$} corresponds to the ratio of geometrical acceptances, reconstruction and selection efficiencies of the signal relative to the normalization decays. 
  
The efficiencies are determined using simulated events that are corrected to account for known differences between data and simulation. A particular challenge is the unknown amplitude composition of the decays under study. The signal decays are simulated with an incoherent sum of resonant and nonresonant dimuon and dihadron components. Samples of background-subtracted and efficiency-corrected \Dhhpmm decays are used to correct the five-dimensional decay model of the corresponding dielectron mode.
In addition, particle-identification, hardware-trigger, and tracking efficiencies as well as the dielectron-mass resolution are corrected using dedicated control channels in data. 

Systematic uncertainties related to the determination of the yields arise due to limited knowledge of the various fit components and are evaluated using pseudoexperiments, where alternative fit models are tested. These include variations of the signal and background shape parameters within the uncertainties determined from the fits to simulation, and a signal shape obtained with a modified electron momentum resolution. In addition, a Bukin distribution is tested as an alternative distribution to represent the signal. Since not all selection criteria can be applied to background simulation because of limited simulated sample sizes, the misidentified background shapes are recomputed using an alternative set of selection criteria. The combinatorial background shape is determined in alternative \dm ranges.  
The fraction of signal decays migrating into other regions of dielectron mass is varied within its uncertainty. Furthermore, fit components are re-evaluated using simulated samples corresponding to different data-taking years. The dominant systematic uncertainty arising from assumptions in the fit is determined by fully neglecting all partially reconstructed backgrounds. The impact of neglecting further misidentified backgrounds is found to be negligible everywhere except in the signal $\rho^0/\omega$ dielectron-mass region, where a contribution from \Dkpeenr decays misreconstructed as the signal are found, and the appropriate systematic uncertainty is computed. 
The statistical uncertainty on the normalization yield leads to a relative systematic uncertainty of $4.8\%$ ($4.5\%$) for \Dppee (\Dkkee). 

Systematic uncertainties affecting the efficiency ratio include residual data-simulation differences and limitations in the data-driven methods used to determine the particle-identification, tracking and trigger efficiencies. 
Uncertainties are evaluated directly on the efficiency ratio to take into account cancellations caused by similarities between the signal and normalization modes. The uncertainty due to finite simulated sample sizes is evaluated using a bootstrapping technique~\cite{efron:1979}. 
The impact of different detector occupancies in data and simulation is evaluated by recomputing the efficiencies using an additional correction accounting for the deviations.
The efficiency ratios are recomputed varying the parameters used to smear the dielectron mass within their statistical uncertainties, and varying the resonant models of the signal decays within their uncertainties. In addition, the efficiency ratios are recomputed after artificially injecting an additional nonresonant component, representing unknown short-distance physics, into the simulated signal distributions at 10\% of the total number of events. Relative variations of the efficiency ratio between approximately 7\% and 14\% are found in the dielectron--mass regions and taken as systematic uncertainties.  The impact of charm hadrons produced in beauty hadron decays is evaluated and found to be negligible. 

To summarize, the biggest systematic uncertainties arise from finite simulated sample sizes, the limited knowledge of the signal-decay resonant structure, and the models used in the fit. 
The total systematic uncertainty depends on the decay mode and dielectron-mass region and ranges between approximately $10\%$ and $70\%$ of the corresponding statistical precision~\cite{supplemental}.

\begin{table}[tb]
\centering
\caption{Branching fractions of (top) \Dppee and (bottom) \Dkkee decays in different ranges of dielectron mass. Shown in comparison on the right are the branching fractions of \Dppmm and \Dkkmm decays in the corresponding dimuon-mass ranges from Ref.~\cite{LHCb-PAPER-2017-019}. The uncertainties are statistical, systematic and due to the limited knowledge of the normalization branching fraction. The reported upper limits correspond to 90\% (95\%) confidence level. The correlations between the various dielectron-mass ranges are reported in the supplemental material \cite{supplemental}.\label{tab:BF}}
\ifthenelse{\boolean{wordcount}}{}{%
\begin{tabular}{lr@{--}lcc}
$m(\ellell)$ region & \multicolumn{2}{c}{[$\mevcc$]} & \multicolumn{2}{c}{$\BF$  [$10^{-7}$]} \\
\hline
& \multicolumn{2}{c}{} & \Dppee & \Dppmm~\cite{LHCb-PAPER-2017-019}\\

Low mass        & 2$m_\mu$ & 525 & $< 4.8\,(5.4)$ & $0.78 \pm 0.19\pm 0.05 \pm 0.08$\\
$\eta$          & 525 & 565  & $<2.3\,(2.7)$ & $< 0.24\,(0.28)$\\
$\rhoz/\omegaz$ & 565 & 950  & $4.5 \pm 1.0\pm 0.7 \pm 0.6$ & $4.06 \pm 0.33\pm 0.21 \pm 0.41$\\
$\phi$          & 950 & 1100 & $3.8 \pm 0.7 \pm 0.4 \pm 0.5$ & $4.54 \pm 0.29\pm 0.25 \pm 0.45$\\
High mass       & \multicolumn{2}{c}{$>1100$} & $<2.0\,(2.2)$& $< 0.28\,(0.33)$ \\
\hline
& \multicolumn{2}{c}{} & \Dkkee & \Dkkmm~\cite{LHCb-PAPER-2017-019}\\
Low mass        &2$m_\mu$ & 525 & $<1.0\,(1.1)$ & $0.26 \pm 0.12\pm 0.02 \pm 0.03$\\
$\eta$          &525 & 565  & $<0.4\,(0.5)$ & $< 0.07\,(0.08)$\\
$\rhoz/\omegaz$ &  \multicolumn{2}{c}{$>565$} & $<2.2\,(2.5)$ & $1.20 \pm 0.23\pm 0.07 \pm 0.12$\\
\hline
\end{tabular}}
\end{table}

Measured branching fractions for \Dppee and \Dkkee decays in regions of $m(\ep\en)$  are reported in Table~\ref{tab:BF} if the statistical significance exceeds three standard deviations, where the first uncertainty accounts for the statistical component, the second for the systematic, and the third corresponds to a $13.7\%$ relative uncertainty on $\BF(\Dkpee)$~\cite{exp:BABAR}. The Feldman--Cousins approach~\cite{PhysRevD.57.3873} is used to report the statistical uncertainties on the branching fractions. Upper limits are derived in all other $m(\ep\en)$ regions using the CL$_{s}$ method 
including the effects due to the systematic uncertainties~\cite{CLs,Junk:1999kv}. For comparison, the known branching fractions of \Dppmm and \Dkkmm decays, taken from Ref~\cite{LHCb-PAPER-2017-019}, are also shown. Integrating over the dielectron-mass regions defined in Table~\ref{tab:BF}, and accounting for correlations~\cite{supplemental}, the total branching fraction for \Dppee decays with $m(\ep\en)$ greater than two times the muon mass is measured to be
\begin{equation*}
     \BF(\Dppeetot) = (13.3 \pm  1.7  \pm  1.7   \pm 1.8 ) \times 10^{-7},
\end{equation*} where the uncertainties are statistical, systematic and due to the limited knowledge of the normalization-mode branching fraction, respectively.
The result is consistent with the SM expectations~\cite{Cappiello,deBoer:2018} and with the branching fraction of \Dppmm of \mbox{$(9.6 \pm 1.2)\times10^{-7}$}~\cite{LHCb-PAPER-2017-019}. No total branching fraction is quoted for \Dkkee decays as no significant signal is observed in any of the dielectron-mass regions.

In summary, a study of the \Dppee and \Dkkee decays is performed in regions of the dielectron mass using \proton\proton collisions collected by the \lhcb experiment at $\sqrt{s}=13\tev$. 
The decay \Dppee is observed for the first time when requiring that the two electrons are consistent with coming from the decay of a $\phi$ or $\rhoz/\omegaz$ mesons. 
No evidence is found for the \Dkkee decay and world-best limits are set on its branching fraction which improves previous best limits by two orders of magnitude. The results are compared to, and found to be consistent with, the branching fractions of the \Dppmm and \Dkkmm decays recently measured by \lhcb, and confirm lepton universality at the current precision.

\newpage
\section*{Acknowledgements}
%
%

 \noindent We express our gratitude to our colleagues in the CERN
 accelerator departments for the excellent performance of the LHC. We
 thank the technical and administrative staff at the LHCb
 institutes.
 We acknowledge support from CERN and from the national agencies:
 ARC (Australia);
 CAPES, CNPq, FAPERJ and FINEP (Brazil); 
 MOST and NSFC (China); 
 CNRS/IN2P3 (France); 
 BMBF, DFG and MPG (Germany); 
 INFN (Italy); 
 NWO (Netherlands); 
 MNiSW and NCN (Poland); 
 MCID/IFA (Romania); 
 MICIU and AEI (Spain);
 SNSF and SER (Switzerland); 
 NASU (Ukraine); 
 STFC (United Kingdom); 
 DOE NP and NSF (USA).
 We acknowledge the computing resources that are provided by ARDC (Australia), 
 CBPF (Brazil),
 CERN, 
 IHEP and LZU (China),
 IN2P3 (France), 
 KIT and DESY (Germany), 
 INFN (Italy), 
 SURF (Netherlands),
 Polish WLCG (Poland),
 IFIN-HH (Romania), 
 PIC (Spain), CSCS (Switzerland), 
 and GridPP (United Kingdom).
 We are indebted to the communities behind the multiple open-source
 software packages on which we depend.
 Individual groups or members have received support from
 Key Research Program of Frontier Sciences of CAS, CAS PIFI, CAS CCEPP, 
 Fundamental Research Funds for the Central Universities,  and Sci.\ \& Tech.\ Program of Guangzhou (China);
 Minciencias (Colombia);
 EPLANET, Marie Sk\l{}odowska-Curie Actions, ERC and NextGenerationEU (European Union);
 A*MIDEX, ANR, IPhU and Labex P2IO, and R\'{e}gion Auvergne-Rh\^{o}ne-Alpes (France);
 Alexander-von-Humboldt Foundation (Germany);
 ICSC (Italy); 
 GVA, XuntaGal, GENCAT, Inditex, InTalent and Prog.~Atracci\'on Talento, CM (Spain);
 Severo Ochoa and Mar\'ia de Maeztu Units of Excellence, GVA, XuntaGal, GENCAT, InTalent-Inditex and Prog.~Atracci\'on Talento CM (Spain);
 SRC (Sweden);
 the Leverhulme Trust, the Royal Society and UKRI (United Kingdom).

\indent
\section*{Data availability}
Data associated to the plots in this publication as well as
in supplementary materials are made available on the
CERN document server in Ref.~\cite{cds_article}.


\newpage
\addcontentsline{toc}{section}{References}
\bibliographystyle{LHCb}
\bibliography{main,standard,LHCb-PAPER,LHCb-CONF,LHCb-DP,LHCb-TDR}

\clearpage
\renewcommand{\thefigure}{S\arabic{figure}}
\renewcommand{\thetable}{S\arabic{table}}
\renewcommand{\theequation}{S\arabic{equation}}

\setcounter{equation}{0}
\setcounter{figure}{0}
\setcounter{table}{0}
\setcounter{page}{1}

\section*{Supplemental material for the Letter ``Search for \Dz meson decays to \ppee and \kkee final states''}

\begin{figure}[h]
\centering
\includegraphics[width=0.45\textwidth]{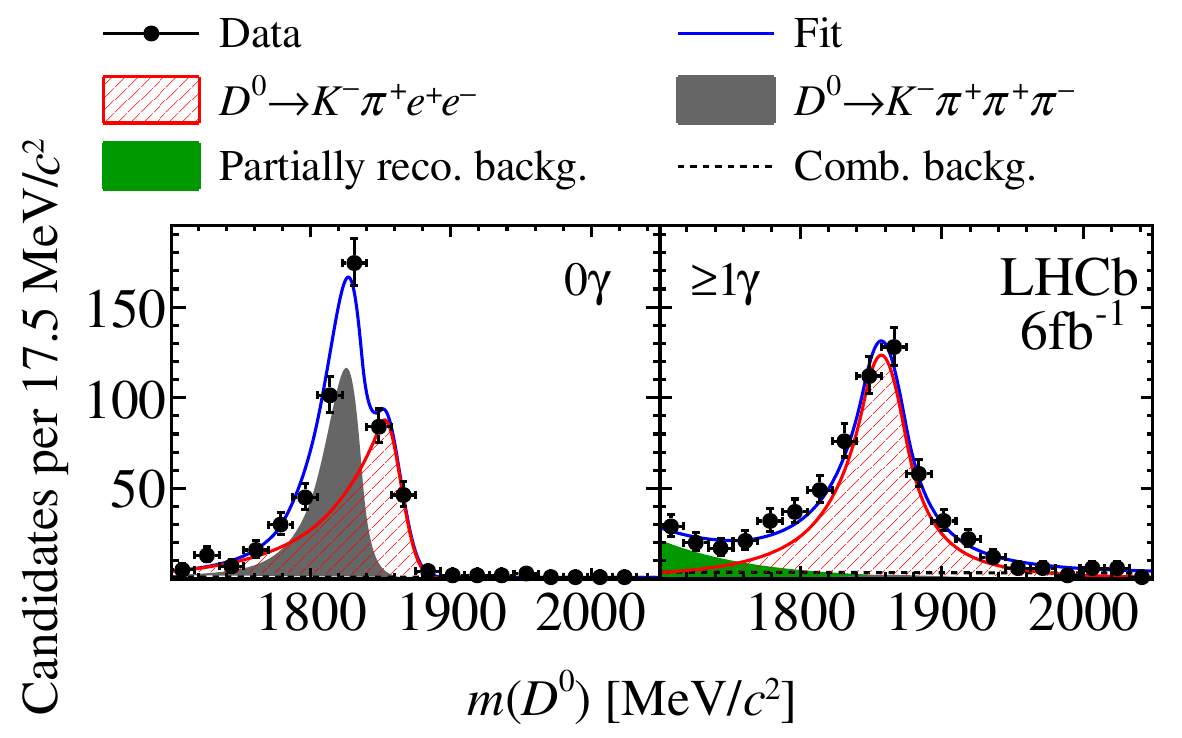}
\includegraphics[width=0.45\textwidth]{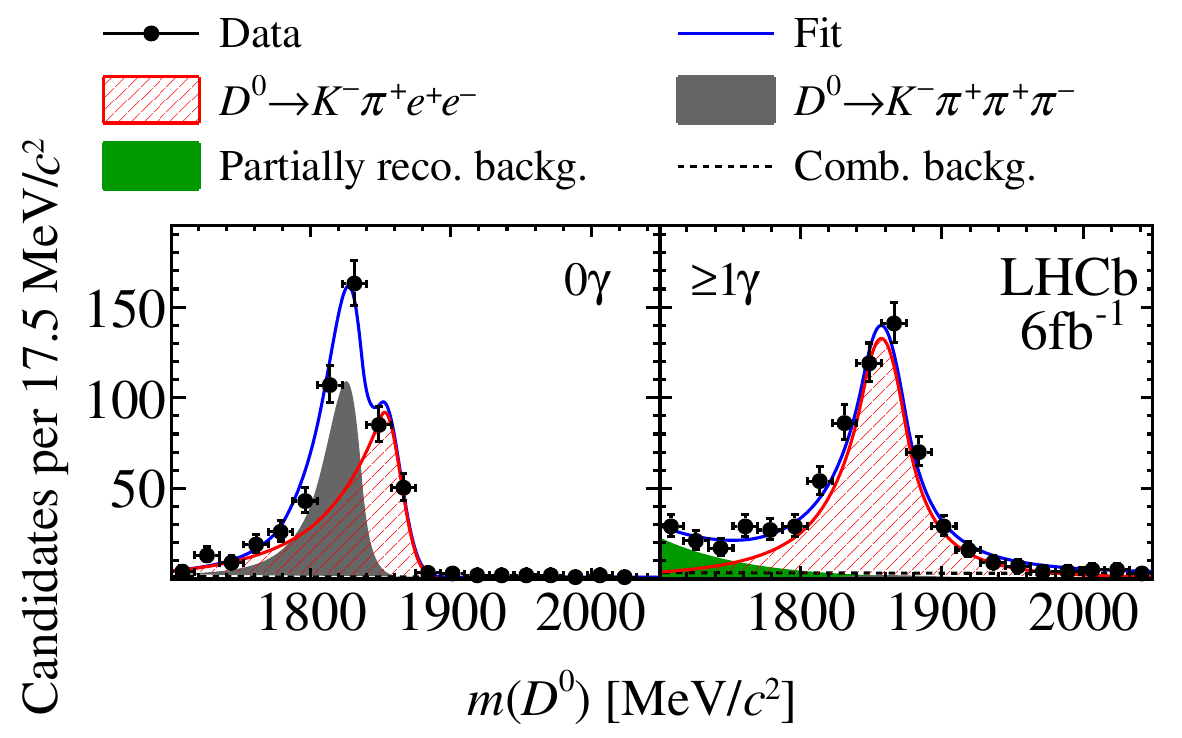}
\caption{Distributions of $m(D^0)$ for selected $D^0 \rightarrow K^-\pi^-[e^+e^-]_{\rho^0/\omega}$ candidates in the dielectron-mass region between 675--875 {Me\kern -0.1em V\!/}$c^2$ split within no-brem and with-brem categories, with the selection optimized for (left) $D^0 \rightarrow \pi^+\pi^-e^+e^-$ and (right) $D^0 \rightarrow K^+K^-e^+e^-$ decays. Fit projections are also shown. \label{fig:massfits-kpiee}}
\end{figure}

The fit results for \Dkpee decays are shown in Fig.~\ref{fig:massfits-kpiee} for candidates selected as normalization sample for measurement of the branching fraction of both \Dppee and \Dkkee decays. The measured branching fractions in dielectron-mass regions for \mbox{\Dppee} and \mbox{\Dkkee} decays are reported in Table~\ref{tab:BF_all_ppee} and Table~\ref{tab:BF_all_KKee}, respectively. A summary of the systematic uncertainties is listed in Table~\ref{tab:systematics}. The correlation matrices for the measurements, including statistical and systematic uncertainties, are given in Tables~\ref{table:corr_ppee} and~\ref{table:corr_KKee}. 

\begin{table}[h]
\centering
\caption{Measured branching fractions of \Dppee decays in regions of the dielectron mass, which are used to compute the total branching fraction. The total uncertainty is reported, including the statistical, systematic and due to the limited knowledge of the normalization mode branching fraction uncertainties.\label{tab:BF_all_ppee}}
\begin{tabular}{lr@{--}lc}
$m(\epem)$ region & \multicolumn{2}{c}{[$\mevcc$]} & $\BF$ [$10^{-7}$]\\
\hline
\multicolumn{4}{c}{\Dppee}\\
Low mass        & 2$m_\mu$ & 525 & $2.8 ^{+1.2}_{-1.1}   $ \\
$\eta$          & 525 & 565  & $1.0 ^{+0.7}_{-0.6} $ \\
$\rhoz/\omegaz$ & 565 & 950  & $4.5 \pm 1.4$\\
$\phi$          & 950 & 1100 & $3.8 \pm 0.9 $ \\
High mass       & \multicolumn{2}{c}{$>1100$} & $1.1 ^{+0.5}_{-0.4} $   \\
\hline
\end{tabular}
\end{table}

\begin{table}[tb]
\centering
\caption{Measured branching fractions of \Dkkee decays in regions of dielectron mass. The total uncertainty is reported, including the statistical, systematic and due to the limited knowledge of the normalization mode branching fraction uncertainties.\label{tab:BF_all_KKee}}
\begin{tabular}{lr@{--}lc}
$m(\epem)$ region & \multicolumn{2}{c}{[$\mevcc$]} & $\BF$ [$10^{-8}$]\\
\hline
\multicolumn{4}{c}{\Dkkee}\\
Low mass        & 2$m_\mu$ & 525 & $0.9 ^{+4.3}_{-1.1} $ \\
$\eta$          & 525 & 565  & $0.1 ^{+1.7}_{-0.2}  $\\
$\rhoz/\omegaz$ & \multicolumn{2}{c}{$>565$}  & $9.9 ^{+6.4}_{-5.1} $\\
\hline
\end{tabular}
\end{table}

\begin{table}[tb]
\centering
\caption{Summary of relative systematic uncertainties on ratio of yields ($\frac{\Delta R_\epsilon}{R_\epsilon}$) and ratio of efficiencies ($\frac{\Delta R_\epsilon}{R_\epsilon}$) for (top) \Dppee and (bottom) \Dkkee decays in different ranges of dielectron mass. The last column shows the sum in quadrature of the contributions. The uncertainty due to the limited knowledge of the normalization mode branching fraction (13.7\%) is not shown in this table.\label{tab:systematics}}
\ifthenelse{\boolean{wordcount}}{}{%
\begin{tabular}{lr@{--}lccc}
$m(\epem)$ region & \multicolumn{2}{c}{[$\mevcc$]} & $\frac{\Delta R_N}{R_N}$[\%] &                                     $\frac{\Delta R_\epsilon}{R_\epsilon}$[\%] & $\frac{\Delta \BF}{\BF}$[\%]\\
\hline\multicolumn{6}{c}{\Dppee}\\

Low mass        & 2$m_\mu$ & 525 & 8.1    & 10.1  & 12.9 \\
$\eta$          & 525 & 565  & 11.7   & 16.9  & 20.6 \\
$\rhoz/\omegaz$ & 565 & 950  & 15.1   & 4.8  & 15.8 \\
$\phi$          & 950 & 1100 & 8.2   & 5.9  & 10.1   \\
High mass       & \multicolumn{2}{c}{$>1100$} & 11.9  & 13.0  & 17.6  \\
\hline
\multicolumn{6}{c}{\Dkkee}\\
Low mass        &2$m_\mu$ & 525 & 9.2   & 8.8  & 12.7  \\
$\eta$          &525 & 565  & 9.8  &  15.4   & 18.3  \\
$\rhoz/\omegaz$ &  \multicolumn{2}{c}{$>565$} & 10.3  &  5.5   & 11.7 \\
\hline
\end{tabular}}
\end{table}

\begin{table}[h]
\centering
\caption{Correlation coefficients related to the statistical and systematic uncertainties of the branching fractions of \Dppee decays in different dilepton mass regions. \label{table:corr_ppee}}
\begin{tabular}{l|lllll}
\hline
$m(\epem)$ region [\mevcc] & 2$m_\mu$--525 & 525--565 & 565--950 & 950--1100 & \textgreater{}1100 \\
\hline
2$m_\mu$--525                     & 1.00    & 0.14    & 0.33    & 0.34     & 0.22               \\
525--565                     &         & 1.00    & 0.25    & 0.25     & 0.17               \\
565--950                     &         &         & 1.00    & 0.54     & 0.37               \\
950--1100                    &         &         &         & 1.00     & 0.37               \\
\textgreater{}1100          &         &         &         &          & 1.00    \\
\hline
\end{tabular}
\end{table}

\begin{table}[h]
\centering
\caption{Correlation coefficients related to the statistical and systematic uncertainties of the branching fractions of \Dkkee decays in different dielectron mass regions. \label{table:corr_KKee}}
\begin{tabular}{l|lll}
\hline
$m(\epem)$ region [\mevcc] & 2$m_\mu$--525 & 525--565 & $>565$  \\
\hline
2$m_\mu$--525                     & 1.00    & 0.05    & 0.02                \\
525--565                     &         & 1.00    & 0.17                   \\
$>565$                    &         &         & 1.00                  \\
\hline
\end{tabular}
\end{table}

\clearpage
\newpage
\centerline
{\large\bf LHCb collaboration}
\begin
{flushleft}
\small
R.~Aaij$^{38}$\lhcborcid{0000-0003-0533-1952},
A.S.W.~Abdelmotteleb$^{57}$\lhcborcid{0000-0001-7905-0542},
C.~Abellan~Beteta$^{51}$,
F.~Abudin{\'e}n$^{57}$\lhcborcid{0000-0002-6737-3528},
T.~Ackernley$^{61}$\lhcborcid{0000-0002-5951-3498},
A. A. ~Adefisoye$^{69}$\lhcborcid{0000-0003-2448-1550},
B.~Adeva$^{47}$\lhcborcid{0000-0001-9756-3712},
M.~Adinolfi$^{55}$\lhcborcid{0000-0002-1326-1264},
P.~Adlarson$^{82}$\lhcborcid{0000-0001-6280-3851},
C.~Agapopoulou$^{14}$\lhcborcid{0000-0002-2368-0147}
% ...

{\footnotesize \it

% ...
$^{14}$Universit{\'e} Paris-Saclay, CNRS/IN2P3, IJCLab, Orsay, France\\
% ...
$^{38}$Nikhef National Institute for Subatomic Physics, Amsterdam, Netherlands\\
% ...
$^{47}$Instituto Galego de F{\'\i}sica de Altas Enerx{\'\i}as (IGFAE), Universidade de Santiago de Compostela, Santiago de Compostela, Spain\\
% ...
$^{51}$Physik-Institut, Universit{\"a}t Z{\"u}rich, Z{\"u}rich, Switzerland\\
% ...
$^{55}$H.H. Wills Physics Laboratory, University of Bristol, Bristol, United Kingdom\\
% ...
$^{57}$Department of Physics, University of Warwick, Coventry, United Kingdom\\
% ...
$^{60}$School of Physics and Astronomy, University of Glasgow, Glasgow, United Kingdom\\
$^{61}$Oliver Lodge Laboratory, University of Liverpool, Liverpool, United Kingdom\\
% ...
$^{69}$Syracuse University, Syracuse, NY, United States\\
% ...
$^{82}$Department of Physics and Astronomy, Uppsala University, Uppsala, Sweden, associated to $^{60}$\\
% ...
}
\end{flushleft}


\end{document}